\documentclass{desyproc}

\begin{document}
%------------------------------------
\title{Soft Diffraction and Forward Multiplicity Measurements with TOTEM}

%for single authors the superscripts are optional
\author{{\slshape Fredrik Oljemark$^1$ (on behalf of the TOTEM collaboration)}\\[1ex]
$^1$Department of Physics, University of Helsinki and Helsinki Institute of Physics, Finland }

% if the proceedings are available online (e.g. at Indico)
% please enter the contribution ID or file_name below for the DOI
%\contribID{32}
%\contribID{smith\_joe}
\contribID{oljemark}

% TO THE CONFERENCE EDITORS: 
% please update the following information      
% before sending the template to the authors
% \confID{800}  % if the conference is on Indico uncomment this line

\acronym{EDS'13} % if you want the Acronym in the page footer uncomment this line

\maketitle

\begin{abstract}
%Place your abstract here. It should not exceed 100 words.
%Please do not modify the style of the paper.  
%In particular, do not change width and height of the text and
%observe the page limits. Please don't use footnotes in the abstract or title.
A summary of recent TOTEM double diffraction and charged particle pseudorapidity density
results is given, and single diffraction results are also discussed.
\end{abstract}

\section{Introduction}

%Write the text here. Write the text here. Write the text here. Write the text here. Write the text here. Write the text here. Write the text here. Write the text here. 
Diffractive scattering represents a unique tool for investigating
the dynamics of strong interactions and proton structure. %These
Diffractive events are dominated by soft processes which cannot be calculated
with perturbative QCD. Various model calculations predict diffractive
cross-sections that are markedly different at the LHC energies~\cite{theory1,theory2,theory3}.

Single diffraction (SD)  is the process in which two colliding hadrons interact
inelastically, one proton staying intact while the other dissociates to a hadronic system. %but only one dissociates into a cluster of particles.
The interaction is mediated by an object with the quantum numbers of the vacuum. Experimentally, SD has a signature of
a final state proton opposite a diffractive system with a rapidity gap in between that is
large compared to random multiplicity fluctuations. 
Rapidity gaps are exponentially suppressed in non-diffractive (ND) events~\cite{rapgap}.
Double diffraction (DD) is a similar process in which both colliding hadrons dissociate into hadronic systems. %clusters of particles.
Here the signature is two forward hadronic systems on opposite sides separated by a rapidity gap.  In Central Diffraction (CD) both protons survive, and
the mediating objects fuse to form a central system separated by rapidity gaps from both protons. 

\subsection{TOTEM}
The TOTEM experiment \cite{totem} is a dedicated experiment to study diffraction, elastic scattering and 
the total cross-section at the LHC. It has three subdetectors placed symmetrically on both sides of
the interaction point: Roman Pot detectors to identify leading protons and T1 and T2 telescopes to
detect charged particles in the forward region. T2 consists of Gas Electron Multiplier chambers that detect charged particles 
with $p_{T}>$40~MeV/c at pseudo-rapidities\footnote{$\eta$=-ln[tan($\theta$/2)] where $\theta$ is the polar angle.} of 5.3$<$$|\eta|$$<$6.5. The T1
telescope consists of Cathode Strip Chambers that measure charged particles with $p_{T}>$100~MeV/c
at 3.1$<$$|\eta|$$<$4.7. The Roman pots (RP) are silicon strip detectors at 220m from the interaction point, housed in
a moveable ``pot'' that can approach the beam very closely.
The RPs measure scattered 
protons : with special LHC optics ($\beta^*=$ 90m) they can detect a proton with any amount
of momentum loss for vertical momentum transfers larger than $|t_y|>$0.005 GeV$^2$. 

The diffractive analyses (DD and SD) use proton-proton collision data that TOTEM took in 2011 at 7 TeV, with $\beta^*=$ 90m,
while the $dN/d\eta$ measurement uses a special run with $\beta^*=$ 3.5m and low pileup, also taken at 7 TeV in 2011.

\section{Double diffraction}
In this measurement~\cite{DD}, the double diffractive cross-section was determined in the forward region. 
The DD events were selected by vetoing T1 tracks and requiring tracks in T2, hence selecting events that have two diffractive systems
with 4.7$<$$|\eta|_{min}$$<$6.5, where $\eta_{min}$ is the minimum pseudorapidy of all primary particles produced in the diffractive system.
Although these events are only about 3$\%$ of the total $\sigma_{DD}$, they provide a pure selection of DD events and the measurement is
an important step towards determining if there is a rich resonance structure in the low mass region~\cite{resonances}. To
probe further, the $\eta_{min}$ range was divided into two halves on each side (at $|\eta|=5.9$), providing four subcategories for the measurement.

First, the raw rate of double diffractive events is estimated: the selected sample
is corrected for trigger efficiency, pile-up and T1 multiplicity, and the amount of background is determined. Second, the
visible cross-section is calculated by correcting the raw rate for acceptance and efficiency to detect particles. Lastly, 
the visible cross-section is corrected so that both diffractive systems have 4.7$<$$|\eta|_{min}$$<$6.5. Three kinds of background were considered 
for the analysis: ND, SD and CD. The data-driven ND and SD background estimation methods
were developed to minimize the model dependence, and the values of the background estimates were calculated iteratively. Since the CD background is 
significantly smaller than the ND and SD ones, its estimate was taken from simulation, using the acceptance and $\sigma_{CD}$=1.3~mb
from Phojet~\cite{phojet}. The results for the DD cross section are shown in Table~\ref{results}.

\begin{table}\footnotesize
\centerline{\begin{tabular}{lccccc}
& ${\rm 6.5>\eta_{min}^+>4.7}$ & ${\rm 5.9>\eta_{min}^+>4.7 }$ & ${\rm 6.5>\eta_{min}^+>5.9}$&${\rm 5.9>\eta_{min}^+>4.7}$&${\rm 6.5>\eta_{min}^+>5.9}$  \\
& ${\rm -6.5<\eta_{min}^-<-4.7}$ & ${\rm -5.9<\eta_{min}^-<-4.7}$&${\rm -6.5<\eta_{min}^-<-5.9}$&${\rm -6.5<\eta_{min}^-<-5.9}$&${\rm -5.9<\eta_{min}^-<-4.7}$  \\
\hline
\hline
TOTEM  & 116$\pm$25 & 65$\pm$20 & 12$\pm$5 & 26$\pm$5 & 27$\pm$5  \\
\hline
Pythia & 159 & 70 & 17 & 36 & 36 \\
Phojet & 101 & 44 & 12 & 23 & 23 \\
\end{tabular}}
\caption{\footnotesize Double diffractive cross-section measurements ($\mu$b) in the forward region. The measurements (that were corrected separately for the different ranges in $\eta_{min}$) are given and compared to Pythia and Phojet predictions. Pythia estimate for total $\sigma_{DD}$=8.1~mb and Phojet estimate $\sigma_{DD}$=3.9~mb.}
\label{results}
\end{table}
%}
%\end{wraptable}

\section{Single diffraction}
In this preliminary measurement SD events (proton + gap + hadronic system) are selected using the RPs and T2. 
We require exactly one proton, with
a rapidity gap in one T2 on the same side as the proton, and T2 tracks on the opposite side. T2 is used as trigger. The 
gap size from the proton to the nearest track in T1 or T2 corresponds to diffractive mass ranges given in Table~\ref{tab:limits}.

The raw rate was first corrected for T2 trigger efficiency and RP acceptance.
Then we subtracted the independent 2-proton background, the pileup of a proton in RP from the beam halo (or $M_{SD}<3$ GeV) with a non-related minimum bias event in T2, and corrected for the probability of the proton producing a shower within the RP station (not reconstructed as a single track). This gave an estimate of the SD signal. 
Migration between rapidity gap categories, beam divergence corrections, and momentum loss resolution effects on the mass spectrum still have to be taken
into account.

The final goal of this analysis is to measure, for each mass bin ($\xi$-bin) in Table~\ref{tab:limits}, the integrated cross section,
the differential spectrum $d\sigma/dt$, and a fitted exponential slope for the spectrum.  If there are any visible effects of low mass
resonances~\cite{resonances}, they will affect what the spectrum looks like at the smallest t-values; the spectrum may turn down, or 
have a dip with respect to the fitted exponential.

\begin{table}%{r}{0.7\textwidth}
\centerline{\begin{tabular}{|l|l|l|l|}
\hline
SD class  & Inelastic telescope configuration & Mass range & Momentum loss ($\xi$) \\
\hline
Low mass & T2 opposite p only (no T1)  & 3.4 -- 8 GeV & $2*10^{-7}$ -- $10^{-6}$ \\
Medium mass & T2 opposite p + T1 opposite p & 8 -- 350 GeV & $10^{-6}$ -- 0.25\%\\
High mass & T2 opposite p + T1 same side as p & 0.35 -- 1.1 TeV & 0.25\% -- 2.5\%\\
%Very high mass & p + T2 both sides & $>$1.1 TeV&$>$2.5\%\\
\hline
\end{tabular}}
\caption{SD classes used}
\label{tab:limits}
\end{table}

\section{Differential charged particle density}

The TOTEM experiment has measured~\cite{dNdEta} the charged particle pseudorapidity density dN$_{\textnormal{ch}}$/d$\eta$ in $pp$ collisions 
at $\sqrt{s} =$ 7 TeV for $5.3<|\eta|<6.4$ in events with at least one charged particle with transverse momentum above 40 MeV/c in this
pseudorapidity range. This extends the analogous measurement performed by the other LHC experiments to the previously unexplored 
forward $\eta$ region. The measurement refers to more than  99\% of non-diffractive processes and to single and double diffractive
processes with diffractive masses above $\sim\,$3.4 GeV, corresponding to about 95\% of the total inelastic cross-section. The
dN$_{\textnormal{ch}}$/d$\eta$ is found to decrease with $|\eta|$, from 3.84 $\pm$ 0.01(stat) $\pm$ 0.37(syst) at
$|\eta| = 5.375$  to 2.38 $\pm$ 0.01(stat) $\pm$ 0.21(syst) at $|\eta| = 6.375$. Several MC generators are
compared to data; none of them are found to fully describe the measurement.

TOTEM  is presently~\cite{Eight} finalizing with CMS an article about a combined measurement at 8 TeV, based on a common run with integrated data taking in 2012.

\begin{figure}[hb]
\centerline{\includegraphics[width=0.45\textwidth]{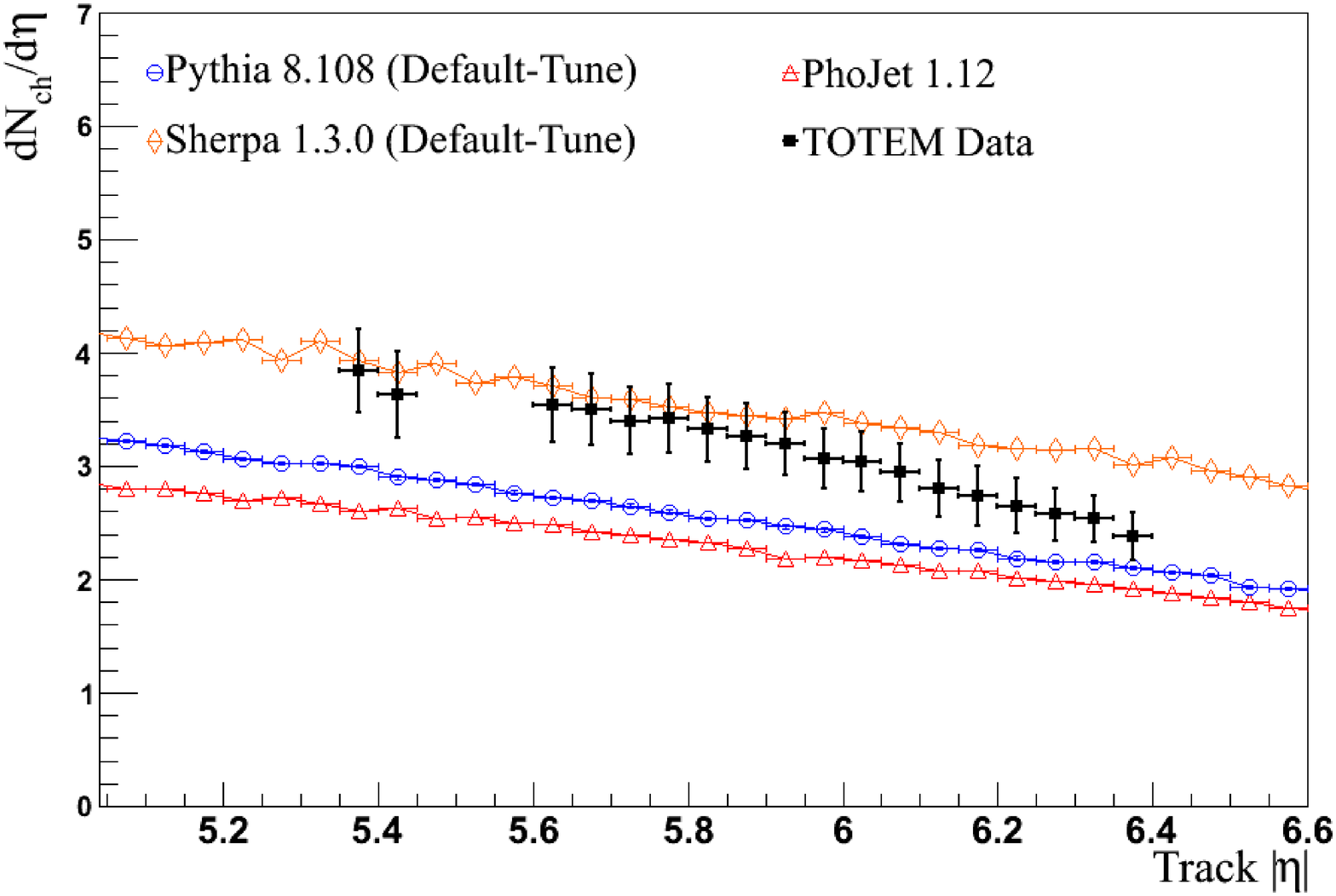}}
\caption{\footnotesize Differential distribution in pseudorapidity. Experimental points (black squares) are the average of the four T2 quarters. Error bars include statistical and systematic errors. Red triangles, blue circles and orange diamonds are: the Phojet, Pythia8 and Sherpa predictions for charged particles with $P_{T}>\,$40 MeV/c in events with $\geq1$ charged particle generated in the $5.3<|\eta|<6.5$ range.}\label{fig:dN}
\end{figure}

%Figure \ref{FigureLabel} shows an example of a figure and related
%caption. This is how you reference an article~\cite{H1}.  

\section{Acknowledgments}

The author acknowledges support from Nylands Nation and Helsinki Institute of Physics.
%To acknowledge funding bodies etc., a special section may be placed
%before the bibliography.

% ****************************************************************************
% BIBLIOGRAPHY AREA
% ****************************************************************************

\begin{footnotesize}
% IF YOU DO NOT USE BIBTEX, USE THE FOLLOWING SAMPLE SCHEME FOR THE REFERENCES
% ----------------------------------------------------------------------------

\end{footnotesize}
\end{document}